\documentclass[prb,eqsecnum,showpacs,amsmath,twocolumn]{revtex4}
\usepackage{graphicx,color}
\usepackage{dcolumn}% Align table columns on decimal point
\usepackage{ascmac}
\usepackage{amsmath,cases}
\usepackage[psamsfonts]{amssymb}
\usepackage{revsymb}
\usepackage{bm}

\newcommand{\ket}{\rangle}
\newcommand{\bra}{\langle}
\renewcommand{\u}{\uparrow}
\renewcommand{\d}{\downarrow}
\newcommand{\nn}{\nonumber}
\newcommand{\Tr}{\mathop{\mathrm{Tr}}\nolimits}

\newcommand{\1}{\openone}

\newcommand{\X}{\text{X}}
\newcommand{\A}{\text{A}}

\begin{document}
%---------- FRONT MATTERS ---------------------------------------
\title{A controlled-{\small{NOT}} gate in a chain of qubits embedded in a spin field-effect transistor and its process tomography}
\author{Makoto Unoki} \email{unoki@hep.phys.waseda.ac.jp}
\affiliation{Department of Physics, Waseda University, Tokyo 169-8555, Japan}
\author{Hiromichi Nakazato}
\affiliation{Department of Physics, Waseda University, Tokyo 169-8555, Japan}
\author{Kazuya Yuasa}
\affiliation{Waseda Institute for Advanced Study, Waseda University, Tokyo 169-8050, Japan}
\author{Kanji Yoh}
\affiliation{Research Center for Integrated Quantum Electronics, Hokkaido University, Sapporo 060-8628, Japan}

\begin{abstract}
We have investigated the realizability of the controlled-\textsc{not} (\textsc{cnot}) gate and characterized the gate operation by quantum process tomography for a chain of qubits, realized by electrons confined in self-assembled quantum dots embedded in the spin field-effect transistor.
We have shown that the \textsc{cnot} gate operation and its process tomography are performable by using the spin exchange interaction and several local qubit rotations within the coherence time of qubits.
Moreover it is shown that when the fluctuation of operation time and the imperfection of polarization of channel electrons are considered as sources of decay of fidelity, at most only 5\% decrease of the \textsc{cnot} process fidelity is expected by the fluctuation of the operation time and its values as high as 0.49 and 0.72 are obtained for the channel spin polarizations of 0.6 and 0.8, respectively.
\end{abstract}
\maketitle

%==========================================================================
%========================= TEXT START =====================================
%==========================================================================
\section{Introduction}
In order to realize quantum information processors, various kinds of physical systems have been proposed and investigated.
In particular, solid-state devices with the quantum bits (qubits) realized by the spins of electrons confined in quantum dots in semiconductors \cite{LV98, F02, HKPTV07} are supposed to be promising in terms of its compatibility with existing semiconductor technology.
Among them, vertically stacked self-assembled InAs quantum dots have an advantage of operational ability at relatively higher temperature of the order of 1 K because of the strong confinement of electrons.\cite{T1}
Recall that an electron gas confined in two dimensions by Schottky electrode operates at most at millikelvin order\cite{F02, HKPTV07}.
Furthermore the vertically stacked quantum dot system has higher scalability than the two dimensional electron gas system.
We have proposed and investigated,\cite{YYN05, KMY07, YONKY09} from both experimental and theoretical aspects, a system of vertically stacked self-assembled InAs dots buried in AlInAs barrier layer adjacent to the channel of a spin field-effect transistor (FET).

In the proposed setup \cite{YYN05, KMY07, YONKY09} illustrated in Fig.\ref{device}, each qubit evolves under the interactions with the neighboring qubits and would be rotated via electric spin resonance (ESR).
Moreover, it is possible to measure repeatedly the spin state of the electron in the edge quantum dot, just above the channel of the FET, by making use of the so-called spin-blockade measurement.
Although we cannot directly access to the other qubits than the one on the edge for the proposed system, we can still perform useful operations on all the qubits.
Initialization and entanglement generation of multiple qubits can be realized via repeated measurements only on the edge qubit.\cite{YYN05}
The state tomography of qubits was discussed in Ref.\,\onlinecite{YONKY09}.
In this paper, we show that a controlled-\textsc{not} (\textsc{cnot}) operation and a characterization of the operation by a quantum process tomography \cite{CN97, PCZ97} are also available in this system.

A \textsc{cnot} gate is one of the most important gate operations because it constitutes a universal set of quantum gates together with singe-qubit rotations.
The \textsc{cnot} gate operation has been carried out for several physical systems including linear optics,\cite{OPWRB03, PFJF03} trapped atoms,\cite{M95, SK03, I10} and solid-state qubits.\cite{YPANT03, PGAM07}
In Ref.\,\onlinecite{H07}, it has been shown that the \textsc{cnot} gate operation can be implemented by using single-qubit rotations and neighboring spin exchange interaction.
We investigate whether the scheme in Ref.\,\onlinecite{H07} can be applied to the proposed system with experimentally reasonable parameters.
Furthermore in order to check that the quantum device operates correctly, it is required to characterize the gate operation. 
In this paper, we characterize the \textsc{cnot} gate operation for the proposed system by a quantum process tomography (QPT) with fluctuations of the operation time and incompleteness of the spin polarization of channel electrons taken into account.
Several QPT experiments have been demonstrated in NMR implementations,\cite{CCL01, W04} optical systems,\cite{MESS03, O04} and in solid-state qubits.\cite{H06, B10}
Notice that we are allowed to measure only the edge qubit of the chain of the qubits. 
Still, we wish to carry out the complete QPT, and we are going to show that it is actually possible.\cite{BMN09}

This article is organized as follows.
In Sec. \ref{structure}, we describe the system of vertically stacked self-assembled InAs quantum dots buried in AlInAs barrier layer adjacent to the channel of a spin FET and some characteristic properties of this system.
In Sec. \ref{sec-CNOT}, we introduce a scheme to realize the \textsc{cnot} operation and discuss a possibility of implementing this scheme to the proposed system with reasonable parameters.
The quantum process tomography of the \textsc{cnot} gate is conducted on the basis of the state tomography scheme \cite{YONKY09} in Sec. \ref{QPT}.
Conclusion and future direction are given in Sec. \ref{conclusion}.

%============================ Section 2 ===============================
\section{Device structure} \label{structure}
%---------------- figure device ------------------
\begin{figure}
\begin{minipage}{\linewidth}
\includegraphics[width=\linewidth]{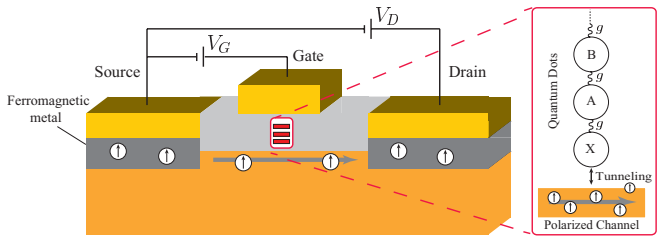}
\end{minipage}
\caption{Schematic view of spin FET embedded with quantum dots.}
\label{device}
\end{figure}
%-------------------------------------------------
The proposed device is illustrated in Fig. \ref{device}.
A series of self-assembled quantum dots are embedded in the FET structure just above the spin channel.
A single electron is confined in each quantum dot and quantum information is encoded on its spin states, $|{\u}\ket$ and $|{\d}\ket$.
Such a situation where only a single electron is stored in each dot is realized by properly adjusting the gate voltage $V_G$.
Each qubit can be rotated via ESR to perform single-qubit operations, and the qubits are made entangled by the interactions between the adjacent qubits.

The FET structure enables us to measure the spin state of the electron confined in the edge quantum dot $X$.
If the channel electrons are all in a definite spin state, say $|{\u}\ket$, the channel current $I_D$ exhibits either peaks or a monotonous increase as a function of the gate voltage $V_G$, depending on the spin state of the electron in dot $X$.
When the spin state of $X$ is $|{\d}\ket_X$, a channel electron enters the edge dot $X$ at a certain gate voltage $V_G$ by the tunneling effect.
On the contrary, when the spin state of $X$ is $|{\u}\ket_X$, it cannot enter the dot due to Pauli's exclusion principle (spin blockade).
Thus one can measure the spin state of the edge qubit by sweeping the gate voltage $V_G$ and exploring the channel current $I_D$.

The feasibility of the present scheme is discussed in Ref.\,\onlinecite{KMY07}, in which the exchange interaction energy $J$ between adjacent qubits and the corresponding time scale $\tau=\pi/J$ are evaluated.
ESR absorption peak separations for quantum dots with a slightly modified compound ratio $x$ of the $\text{In}_{1-x}\text{Ga}_x\text{As}$, which is necessary to a selective access to each qubit, are also investigated.
For a selective spin flipping in stacked quantum dots, difference of more than 10\% in the composition ratio of adjacent dots is needed.
The spin-flipping time is estimated to be 50 ps.
The modulation of the channel current by a single-electron charging in a quantum dot adjacent to the channel is demonstrated in a trial structure with a single layer of quantum dot with unpolarized channel current.
A spin decoherence time of a few ns has been reported for InAs self-assembled quantum dots.\cite{WASM05}
The electrical injection of spin-polarized electrons into semiconductors is investigated and demonstrated recently \cite{Z01, J05, R08} and the spin injection from ferromagnetic metals is demonstrated for several systems.\cite{Z01, J05}
In the case of Heusler alloys as a spin injector, a spin-injection efficiency of at least 50\% is already achieved.\cite{R08}
The spin injection with the high efficiency of 80\% is also realized in the diluted magnetic semiconductor device.\cite{KKOMO06}
Since hole conduction is used in the case of diluted magnetic semiconductor, the whole device must be changed into a hole system but it is expected that such efficient spin injection should be attained in the proposed system.

%============================ Section 3 ===============================
\section{Realization of \textsc{cnot} gate} \label{sec-CNOT}
In this study, we consider a two-qubit system, i.e., only qubits X and A are considered.
The \textsc{cnot} gate is represented by the following unitary transformation 
\begin{equation}
U_\textsc{cnot}=|{\u}\ket_X\bra{\u}|\otimes\1_A+|{\d}\ket_X\bra{\d}|\otimes\sigma_x^A. \label{eq:cnot}
\end{equation}
Here qubit $X$ and $A$ are the control and target qubit, respectively and $|{\u}({\d})\ket_X$ is the eigenstate corresponding to eigenvalue $1(-1)$ of $\sigma_z^X$.

A general prescription to construct this operation by using of spin-spin exchange interaction
\begin{equation}
H=g\bm{\sigma}^X\cdot\bm{\sigma}^A
\end{equation}
and single-qubit rotations is proposed by Hill.\cite{H07}
It is however not obvious whether this method is realizable in each physical system or not.
Therefore it is important to investigate the feasibility of a \textsc{cnot} gate in the proposed system. 

The prescription proposed by Hill \cite{H07} is as follows.
Firstly, we perform the so-called term isolation.
The isolation of $\sigma_z^{X}\sigma_z^{A}$ term can be achieved by the following sequence,
\begin{equation}
e^{-igt\sigma_z^X\sigma_z^A}=R^{(X)}_z(\pi)e^{-iHt/2}R^{(X)}_z(\pi)e^{-iHt/2}, \label{term-iso}
\end{equation}
where $R^{({Q})}_z(\theta)=e^{-i\theta\sigma_z^Q/2}$ is a single-qubit rotation of an angle $\theta$ around the $z$ axis of qubit $Q$ ($Q=X$ or $A$) and an overall phase is ignored.
Then we can realize (\ref{eq:cnot}) by using the isolated $\sigma_z^\X\sigma_z^\A$ interaction (\ref{term-iso}) and some single-qubit operations,
\begin{equation}
U_\textsc{cnot}=H^{(A)}R^{(X)}_z\left(\frac{\pi}{2}\right)R^{(A)}_z\left(\frac{\pi}{2}\right)e^{-i\frac{3\pi}{4}\sigma_z^X\sigma_z^A}H^{(A)}, \label{CNOT}
\end{equation}
where $H^{(A)}$ stands for the Hadamard gate on qubit $A$.

In the proposed device, the strength of the exchange interaction $g$ is typically of the order of 0.01-1 meV for stacked self-assembled quantum dots and so the typical operation time $\tau=\frac{\pi}{4g}$ becomes 0.5-50 ps.
This is small enough compared to the spin-decoherence time of self-assembled quantum dots ($\sim$ a few ns) and comparable to the operation time of single-qubit spin rotation, implying that the \textsc{cnot} gate operation can be realized for the proposed system within the coherence time.

%============================ Section 4 ===============================
\section{Process tomography of \textsc{cnot} gate} \label{QPT}
To characterize the \textsc{cnot} gate operation, we perform the quantum process tomography (QPT).
The idea of QPT is to determine a completely positive (CP) map, which represents the process acting on an arbitrary input state $\rho$:
\begin{equation}
\mathcal{E}(\rho)=\sum_i A_i\rho A_i^\dagger, \label{CP}
\end{equation}
where $A_i$ are Kraus operators and satisfy the condition $\sum_i A_i^\dagger A_i=\1$.
If we expand $A_i$ in terms of a basis for operators acting on $\rho$, $\{E_{mn}\}_{m,n=1}^4$, the CP map $\mathcal{E}$ can be rewritten as
\begin{equation}
\mathcal{E}(\rho)=\sum_{m,n,k,l}\chi_{mn,kl}E_{mk}\rho E_{nl}^\dagger,
\end{equation}
where $\chi_{mn,kl}=\sum_i\Tr[E_{mk}^\dagger A_i]\Tr[E_{nl}^\dagger A_i]^*$.
The matrix $\chi=\{\chi_{mn,kl}\}_{m,n,k,l=1}^4$ is called a process matrix.
If all elements of $\chi$ are known, one can obtain output state of CP map (\ref{CP}) for any input state.
When we choose the basis operators $E_{mn}=|m\ket\bra n|$ in terms of the basis vectors $\{|m\ket\}_{m=1}^4$ in Hilbert space, $\chi_{mn,kl}$ is represented by $\chi_{mn,kl}=\Tr[E_{mn}^\dagger \mathcal{E}(E_{kl})]=\bra m|\big\{\mathcal{E}(|k\ket\bra l|)\big\}|n\ket$.
We can determine the process matrix elements as the output matrix elements of the CP map for the specific 16 inputs $\{E_{kl}\}$.

In order to obtain these process matrix elements $\chi_{mn,kl}$ we perform the following three steps of operation.

\noindent
Step 1: preparation of initial states.
In this paper, we choose the following bases of Hilbert space
\begin{eqnarray}
|1\ket&=&|{\u\u}\ket_{XA}, \\
|2\ket&=&|{\u\d}\ket_{XA}, \\
|3\ket&=&|{\d\u}\ket_{XA}, \\
|4\ket&=&|{\d\d}\ket_{XA}.
\end{eqnarray}
Note that $E_{mn}=|m\ket\bra n|$ is not a quantum state for $m\neq n$ since it is not Hermitian, but it can be reconstructed as a linear combination of Hermitian input states $|\pm; mn\ket$
\begin{eqnarray}
E_{mn}&=&|+; mn\ket\bra +; mn|+i|-; mn\ket\bra -; mn| \nn\\
& &{}-\frac{1+i}{2}(|m\ket\bra m|+|n\ket\bra n|), \label{off-diagonal}
\end{eqnarray}
where $|+; mn\ket=\frac{1}{\sqrt{2}}(|m\ket+|n\ket)$, $|-; mn\ket=\frac{1}{\sqrt{2}}(|m\ket+i|n\ket)$.
The 16 input states we must prepare are thus $\{|m\ket|\ m=1,2,3,4\}$ and $\{|\pm; mn\ket|\ m,n=1,2,3,4,\ m<n\}$.

\noindent
Step 2: \textsc{cnot} gate operation.
For the input states $\rho_0=|m\ket\bra m|$, $|\pm; mn\ket\bra\pm; mn|$, we perform the \textsc{cnot} operation described in the previous section.
In this step we take into account of the effect of fluctuation of the operation time $\tau$.
We assume that the operation time is not fixed at $\tau={3\pi\over4g}\equiv\tau_0$, but is fluctuated around it with a Gaussian distribution
\begin{equation}
P(\tau)d\tau=\frac{1}{\sqrt{2\pi(\Delta\tau)^2}}e^{-\frac{(\tau-\tau_0)^2}{2(\Delta\tau)^2}}d\tau, \label{eq: gauss}
\end{equation} 
where $\bra\tau\ket=\tau_0$ and $\sqrt{\bra(\tau-\bra\tau\ket)^2\ket}=\Delta\tau$ are the mean value and the dispersion, respectively and the output state is given by the ensemble average over this distribution.
The output states are not described by (\ref{CNOT}) as $\mathcal{E}_\text{ideal}(\rho_0)=(U_\textsc{cnot})\rho_0 (U_\textsc{cnot})^\dagger$ but by another CP map $\mathcal{E}(\rho_0)=\langle(U_\textsc{cnot}^\prime(\tau))\rho_0 (U_\textsc{cnot}^\prime(\tau))^\dagger\rangle$, where $U_\textsc{cnot}^\prime(\tau)\equiv H^{(A)}R^{(X)}_z\left(\frac{\pi}{2}\right)R^{(A)}_z\left(\frac{\pi}{2}\right)e^{-ig\tau\sigma_z^X\sigma_z^A}H^{(A)}$.

\noindent
Step 3: state tomography of output states.
A state tomography scheme for the system described above is proposed in Ref.\,\onlinecite{YONKY09}.
For this system the state of the edge qubit can be measured repeatedly but the state of the other qubit than the edge qubit cannot be measured directly.
The idea of Ref.\,\onlinecite{YONKY09} is that one obtains the information about this qubit by using the entangling dynamics and the collapse of the state by the measurement on the edge qubit.
If we perform a unitary operation $U$ on the state $\rho$ and measure the edge qubit to confirm that it is in the state $|{\u}\ket_X$, the probability $p$ of getting this result is given by $p=\Tr[P_{\u}U\rho U^\dagger P_{\u}]=\Tr[(U^\dagger P_{\u}P_{\u}U)\rho]$ where $P_{\u}=|{\u}\ket_X\bra{\u}|$ is the projection operator.
This probability can be interpreted as the expectation value of a Hermitian operator $(U^\dagger P_{\u}P_{\u}U)$ in the state $\rho$.
Therefore, by suitably arranging the unitary operations and measurements, we can obtain a set of expectation values of linearly independent operators and reconstruct all the elements of the state.
It was shown \cite{YONKY09} that 15 linearly independent sequences of operations are sufficient to reconstruct a two-qubit state.
Every sequence consists of measurement of the edge qubit $P_{\u(\d)}$, time evolution $U(\tau)=e^{-iH\tau}$, and global rotation $R_i(\theta)=R_i^{(X)}(\theta)R_i^{(A)}(\theta)$, where $R_i^{(Q)}(\theta)=e^{-i\theta\sigma_i^{Q}/2}$ is a local rotation of qubit $Q(=X, A)$ by an angle $\theta$ around the $i(=x, y)$ axis.
Details of sequences are listed in Table I of Ref.\,\onlinecite{YONKY09}.

We consider two types of sources of experimental error in step 3.
The first one is an imperfection of the spin polarization of channel electrons in the measurement process.
This makes the fidelity of spin-blockade measurement degrade.
When the degree of polarization of the channel electrons is $r$ ($0\le r\le1$), the measurement of the spin state of $X$ is not represented by the pure projection $\rho\to P_\u\rho P_\u$ but by 
\begin{equation}
\rho\to\frac{1+r}{2}P_\u\rho P_\u+\frac{1-r}{2}P_\d\rho P_\d,
\end{equation}
where $P_{\u(\d)}=|{\u}({\d})\ket_X\bra{\u}({\d})|$ is the projection operator onto $|{\u}({\d})\ket_X$ state.
This means that when we perform a measurement $X$ to project it to $|{\u}\ket_X$ we obtain the correct result only with a probability $\frac{1+r}{2}$ and then, the fidelity of measurement is degraded.
The second source of error is a fluctuation of the operation time $\tau$.
This can be dealt with in the same way as in step 2.

According to the recipes in Ref.\,\onlinecite{YONKY09} (including small modifications), we can obtain the matrix elements of the state $\mathcal{E}(\rho_0)$, which includes the effects of imperfect polarization $r$ and fluctuations of operation times.
Finally, we convert these results to the process matrix elements $\chi_{mn,kl}(r,\Delta\tau)=\bra m|\mathcal{E}(|k\ket\bra l|)|n\ket=\Tr[E_{mn}^\dagger\mathcal{E}(E_{kl})]$.

For example, in order to obtain a matrix element $\chi_{11,11}(r, \Delta\tau)=\bra 1|\mathcal{E}(|1\ket\bra 1|)|1\ket$, firstly we calculate the result of \textsc{cnot} gate operation $\mathcal{E}$ for the input state $|1\ket\bra1|$.
\begin{align}
\mathcal{E}(|1\ket\bra1|)=&\frac{1}{8}\{(1+d)(3+d)|1\ket\bra 1|+(1-d)(3-d)|2\ket\bra2| \nn\\
&\phantom{\frac{1}{8}\{}
 +(1-d^2)(|3\ket\bra3|+|4\ket\bra4|+|1\ket\bra2|+h.c.)\}, \label{eq: state_11}
\end{align}
where $d=e^{-2(g\Delta\tau)^2}$.
Next we consider the following sequence of operation for the tomography of this state.
\begin{equation}
P_\u \to U(\tau)=e^{-iH\tau} \to P_\u,
\end{equation}
where $\tau$ is fluctuated around the expectation value $\tau_0=\frac{\pi}{4g}$ with the Gaussian distribution (\ref{eq: gauss}).
When the polarization is $r$ the success probability of this sequence is given by
\begin{equation}
p^{\prime(1)}_\u=\frac{1}{4}\{(1+r)^2 p^{(1)}_\u+(1-r^2)(p^{(1)}_\d+p^{(1e)}_\u)+(1-r)^2p^{(1e)}_\d\}, \label{eq: prob1}
\end{equation}
where
\begin{align}
p^{(1)}_{\u(\d)}&=\langle \Tr[(P_{\u(\d)} U(\tau) P_\u)\rho(P_{\u(\d)} U(\tau) P_\u)^\dagger]\rangle, \\
p^{(1e)}_{\u(\d)}&=\langle \Tr[(P_{\u(\d)} U(\tau) P_\d)\rho(P_{\u(\d)} U(\tau) P_\d)^\dagger]\rangle.
\end{align}
This probability for general input state $\rho$ is easily calculated and the result is
\begin{align}
p^{\prime(1)}_\u=&\frac{1}{4}\{(1+r)^2\rho_{11}+(1+r)(1-d^4 r)\rho_{22} \nn\\
&\phantom{\frac{1}{4}\{}
 +(1-r)(1+d^4 r)\rho_{33}+(1-r)^2\rho_{44}\}, \label{eq: prob2}
\end{align}
where $\rho_{ij}=\bra i|\rho|j\ket$.
This probability corresponds to the density matrix element $\rho_{11}$.
In fact for ideal situation, i.e. $r=1$ and $\Delta\tau=0$, the probability (\ref{eq: prob2}) becomes $\rho_{11}$.
Because the process matrix element $\chi_{11, 11}(r, \Delta\tau)=\bra 1|\mathcal{E}(|1\ket\bra1|)|1\ket$ is obtained as the element $\rho_{11}$ of $\rho=\mathcal{E}(|1\ket\bra1|)$, we achieve the following expression by substituting (\ref{eq: state_11}) into (\ref{eq: prob2}),
\begin{align}
\chi_{11, 11}(r, \Delta\tau)=&\frac{1}{16}[4+\{2(1+d)^2+(1-d^4)(1-d)^2\}r \nn\\
&\phantom{\frac{1}{16}[}
 +2\{1+d-d^4(1-d)\}r^2].
\end{align}
Other elements $\chi_{mn,kl}(r, \Delta\tau)$ can be obtained in similar ways with appropriate initial states and sequences of operations.

%--------------------------- figure (process matrix) --------------------
\begin{figure*}%[tbp]
\begin{minipage}{\linewidth}
\includegraphics[width=.35\linewidth]{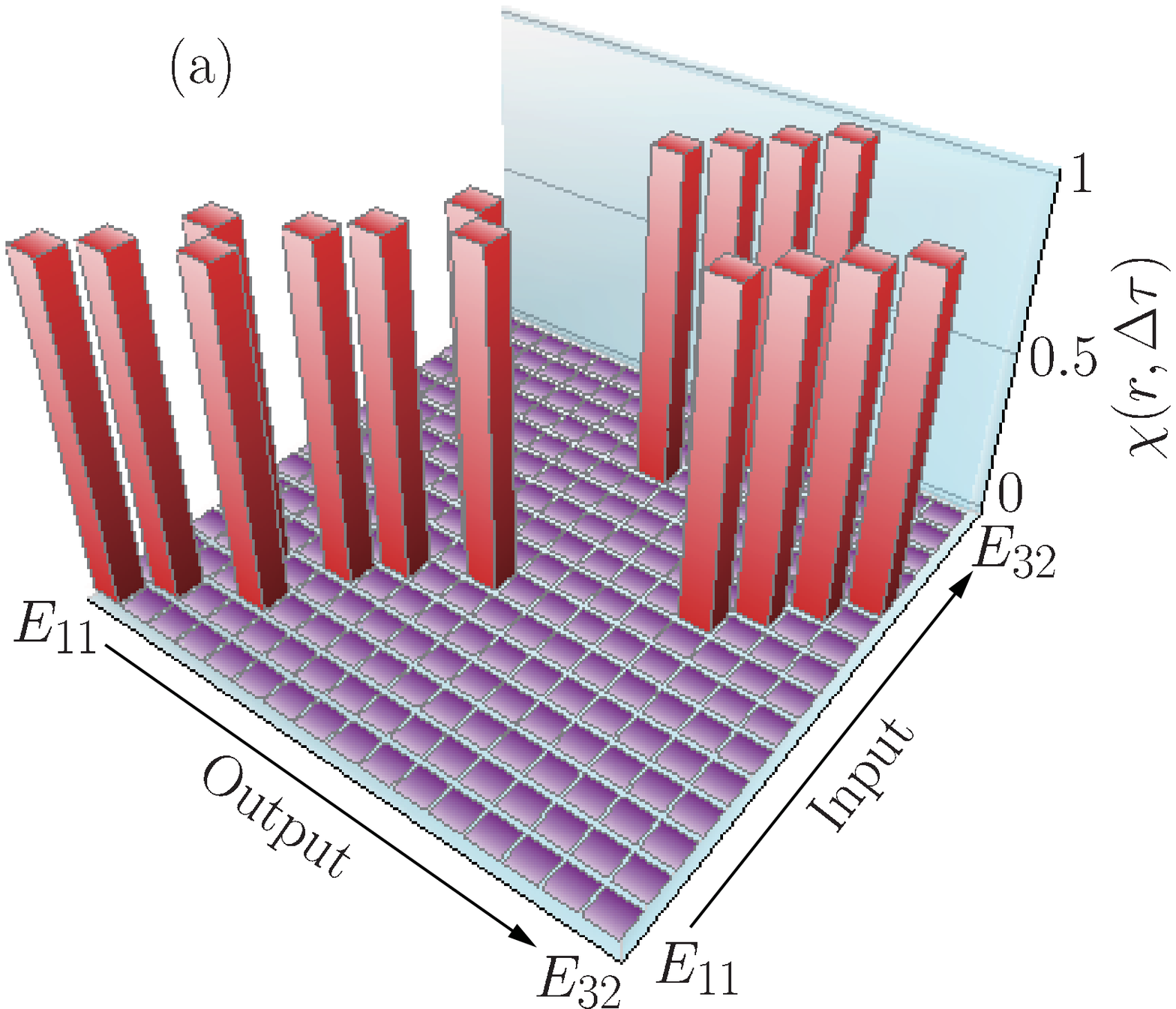}\\
\includegraphics[width=.35\linewidth]{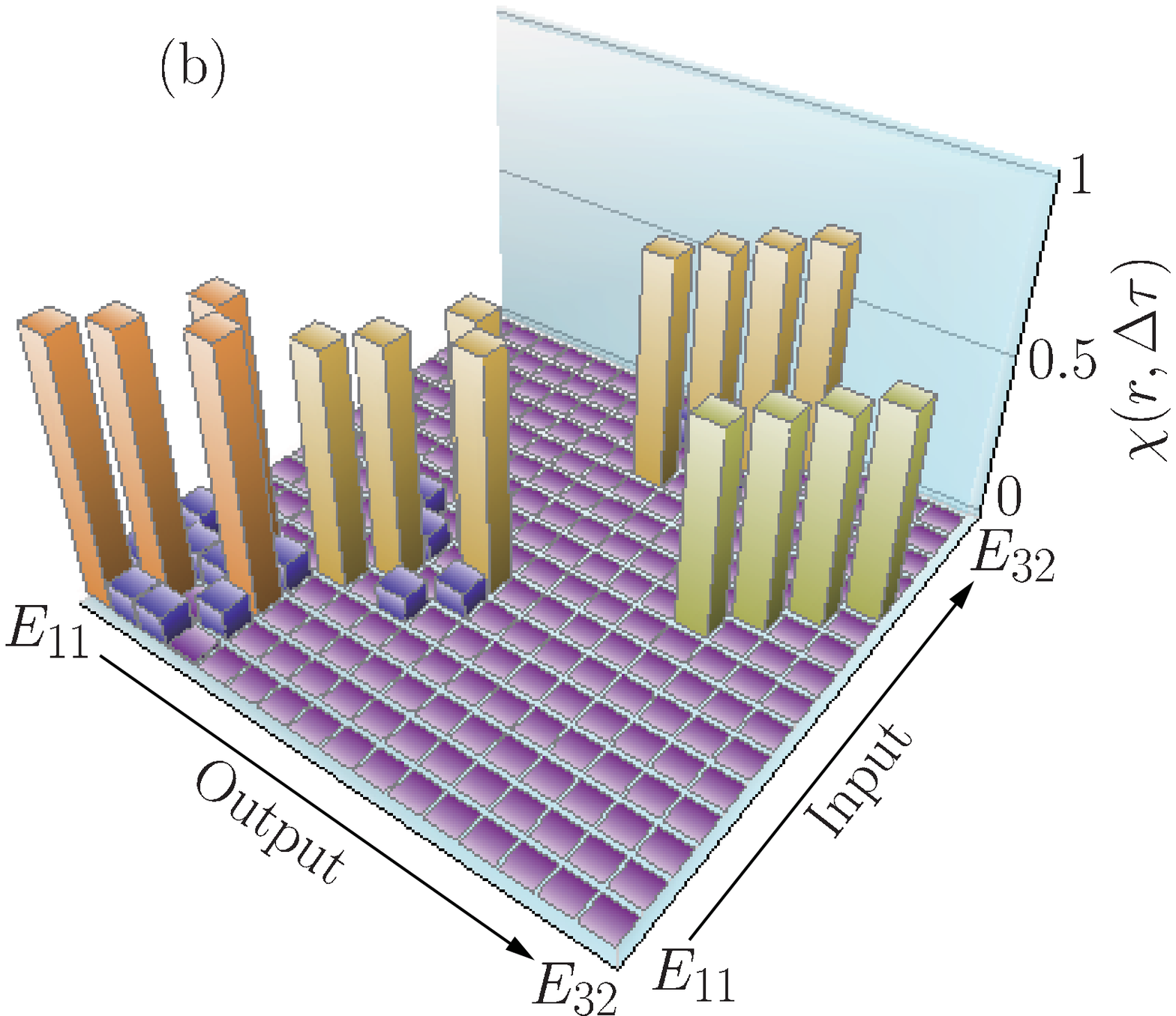}\hspace*{6mm}
\includegraphics[width=.35\linewidth]{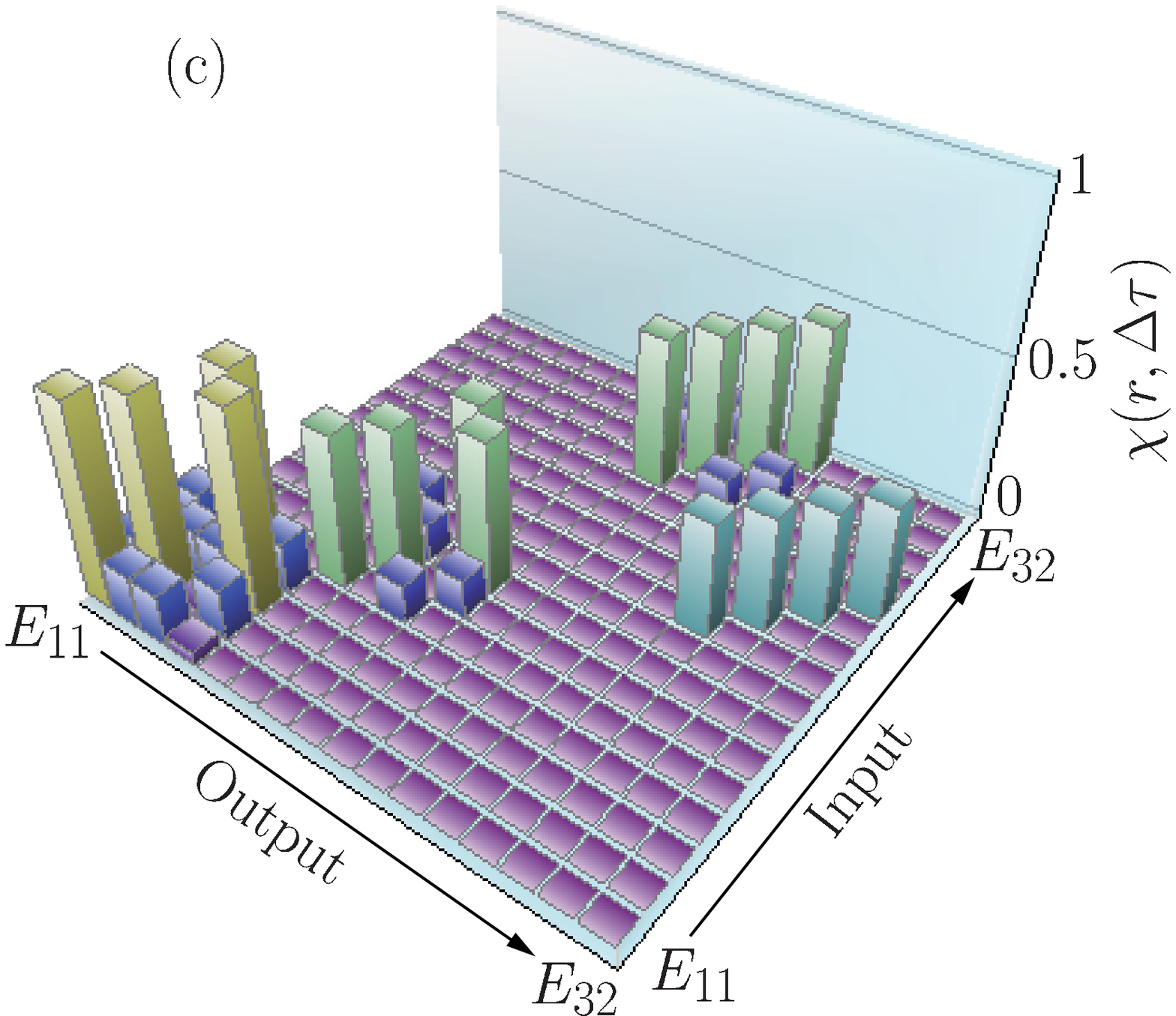}
\end{minipage}
\caption{Process matrix of \textsc{cnot} gate. No fluctuation on the operation time is considered here, i.e., $\Delta\tau=0$. (a) Ideal case with the perfect spin polarization in the channel electrons ($r=1$), and more realistic cases with degraded polarizations (b) $r=0.8$ and (c) $r=0.6$. Bases of input and output operators are orderd as $E_{11}, E_{22}, E_{33}, E_{44}, E_{12}, E_{21}, E_{34}, E_{43}, E_{13}, E_{31}$, $E_{24}, E_{42}, E_{14}, E_{41}, E_{23}, E_{32}.$
}
\label{mat_CNOT}
\end{figure*}
%-----------------------------------------------------------------------

When we choose the bases as $\{ E_{11}, E_{22}, E_{33}, E_{44}| E_{12}, E_{21}, E_{34}, E_{43}| E_{13}, E_{31}, E_{24}, E_{42}|$ $E_{14}, E_{41}, E_{23}, E_{32} \}$, the process matrix $\chi(r,\Delta\tau)$ is given by
\begin{equation}
\chi(r,\Delta\tau)=\frac{1}{4}\left(
\begin{array}{c|c|c|c}
M_1 & M_7 & 0 & 0 \\
\hline
M_8 & M_2 & 0 & 0 \\
\hline
M_9 & M_{10} & M_5 & M_3 \\
\hline
M_{11} & 0 & M_4 & M_6
\end{array}
\right),
\end{equation}
where
\begin{widetext}
\begin{alignat}{2}
M_1&=
	\begin{pmatrix}
	\alpha^{(1)}_{++} & \alpha^{(1)}_{+-} & \alpha^{(1)}_{-+} & \alpha^{(1)}_{--} \\
	\alpha^{(2)}_{+-} & \alpha^{(2)}_{++} & \alpha^{(2)}_{--} & \alpha^{(2)}_{-+} \\
	\alpha^{(2)}_{--} & \alpha^{(2)}_{-+} & \alpha^{(2)}_{+-} & \alpha^{(2)}_{++} \\
	\alpha^{(1)}_{-+} & \alpha^{(1)}_{--} & \alpha^{(1)}_{++} & \alpha^{(1)}_{+-}
	\end{pmatrix},\quad
	&&M_2=2c_+r
	\begin{pmatrix}
	\alpha^{(3)}_{++} & \alpha^{(3)}_{+-} & \alpha^{(3)}_{--} & \alpha^{(3)}_{-+} \\
	\alpha^{(3)}_{+-} & \alpha^{(3)}_{++} & \alpha^{(3)}_{-+} & \alpha^{(3)}_{--} \\
	\alpha^{(3)}_{-+} & \alpha^{(3)}_{--} & \alpha^{(3)}_{+-} & \alpha^{(3)}_{++} \\
	\alpha^{(3)}_{--} & \alpha^{(3)}_{-+} & \alpha^{(3)}_{++} & \alpha^{(3)}_{+-}
	\end{pmatrix},\\
M_3&=c_+^2r
	\begin{pmatrix}
	\alpha^{(4)}_+ & b_-r & \alpha^{(4)}_- & -b_-r \\
	b_-r & \alpha^{(4)}_+ & -b_-r & \alpha^{(4)}_- \\
	\alpha^{(4)}_- & -b_-r & \alpha^{(4)}_+ & b_-r \\
	-b_-r & \alpha^{(4)}_- & b_-r & \alpha^{(4)}_+
	\end{pmatrix},\quad
	&&M_4=c_+r^2
	\begin{pmatrix}
	\alpha^{(5)}_+ & b_- & \alpha^{(5)}_- & -b_- \\
	b_- & \alpha^{(5)}_+ & -b_- & \alpha^{(5)}_- \\
	\alpha^{(5)}_- & -b_- & \alpha^{(5)}_+ & b_- \\
	-b_- & \alpha^{(5)}_- & b_- & \alpha^{(5)}_+
	\end{pmatrix}, \\
M_5&=b_-c_+^2r^2
	\begin{pmatrix}
	-1 & 1 & 1 & -1 \\
	1 & -1 & -1 & 1 \\
	1 & -1 & -1 & 1 \\
	-1 & 1 & 1 & -1
	\end{pmatrix},\quad
	&&M_6=-b_-r
	\begin{pmatrix}
	\alpha^{(6)}_+ & (\alpha^{(6)}_-)^* & \alpha^{(6)}_- & (\alpha^{(6)}_+)^* \\
	\alpha^{(6)}_- & (\alpha^{(6)}_+)^* & \alpha^{(6)}_+ & (\alpha^{(6)}_-)^* \\
	(\alpha^{(6)}_-)^* & \alpha^{(6)}_+ & (\alpha^{(6)}_+)^* & \alpha^{(6)}_- \\
	(\alpha^{(6)}_+)^* & \alpha^{(6)}_- & (\alpha^{(6)}_-)^* & \alpha^{(6)}_+
	\end{pmatrix},
\end{alignat}
\begin{equation}
M_7=b_-r
	\begin{pmatrix}
	1+c_- & 1+c_- & 1+c_- & 1+c_- \\
	c_+ & c_+ & c_+ & c_+ \\
	-c_+ & -c_+ & -c_+ & -c_+ \\
	-(1+c_-) & -(1+c_-) & -(1+c_-) & -(1+c_-)
	\end{pmatrix},
\end{equation}
\begin{equation}
M_8=b_-c_+r
	\begin{pmatrix}
	1 & 1 & -1 & -1 \\
	1 & 1 & -1 & -1 \\
	1 & 1 & -1 & -1 \\
	1 & 1 & -1 & -1
	\end{pmatrix},\quad
M_9=c_- M_8,\quad
M_{10}=c_- M_2,\quad
M_{11}=2c_-r^2
	\begin{pmatrix}
	0 & 0 & 0 & 0 \\
	0 & 0 & 0 & 0 \\
	1 & 1 & 1 & 1 \\
	1 & 1 & 1 & 1
	\end{pmatrix},
\end{equation}
\end{widetext}
with
\begin{alignat}{2}
\alpha^{(1)}_{+\pm}&=1+2\beta^{(1)}_\pm r+\beta^{(3)}_\pm r^2,\quad
&&\alpha^{(1)}_{-\pm}=1-2\beta^{(1)}_\pm r+\beta^{(3)}_\pm r^2,\nonumber\\
\alpha^{(2)}_{+\pm}&=1+2\beta^{(2)}_\pm r-\beta^{(3)}_\mp r^2,\quad
&&\alpha^{(2)}_{-\pm}=1-2\beta^{(2)}_\pm r-\beta^{(3)}_\mp r^2,\nonumber\\
\alpha^{(3)}_{+\pm}&=a_\pm(a_\pm+r),\quad
&&\alpha^{(3)}_{-\pm}=a_\pm(a_\pm-r),\nonumber\\
\alpha^{(4)}_\pm&=2d\pm(1+b_+)r,\quad
&&\alpha^{(5)}_\pm=\pm2a_\pm(1+a_\pm),\nonumber\\
\alpha^{(6)}_\pm&=2(1+i)db_+\pm c_+r,\quad
&&\beta^{(1)}_\pm=a_\pm^2+c_-a_\mp^2,\nonumber\\
\beta^{(2)}_\pm&=c_+a_\pm^2,\quad
&&\beta^{(3)}_\pm=a_\pm-d^4a_\mp,\nonumber\\
a_\pm&=\frac{1}{2}(1\pm d),\quad
&&b_\pm=\frac{1}{2}(1\pm d^2),\nonumber\\
c_\pm&=\frac{1}{2}(1\pm d^4),\quad
&&d=e^{-2(g\Delta\tau)^2}.
\end{alignat}

Figures \ref{mat_CNOT} (a), (b) and (c) show the elements of the process matrix in the case of $\Delta\tau=0$, for $r=1$ (ideal case), $0.8$ and $0.6$, respectively.
Since the effect of fluctuation of operation time $\Delta\tau$ is found to be small, $\Delta\tau\neq0$ case is not shown here.
In fact the effect of the fluctuation of operation time appears through $d=e^{-2(g\Delta\tau)^2}$ and the value of $g\Delta\tau$ is at most 0.1, which is corresponding to the case of $\Delta\tau/\tau \sim 1$ \%, then the deviation of $d$ from the ideal value 1 is at most a few percent.
On the contrary, most of the matrix elements contain terms that are degraded due to the imperfection of spin polarization in proportion to the polarization $r$.

%----------------------- figure (process fidelity) ------------------
\begin{figure}
\includegraphics[width=.9\linewidth]{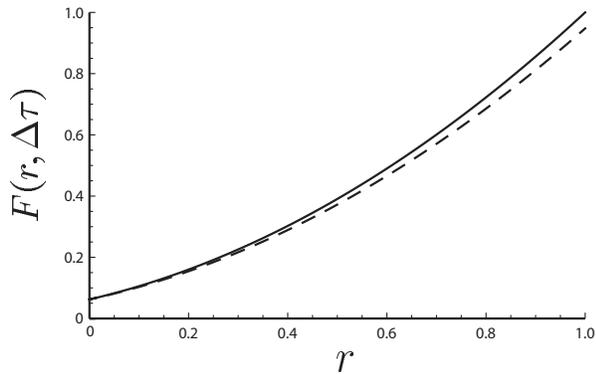}
\caption{Process fidelity of \textsc{cnot} gate as a function of polarization $r$. 
Solid line stands for $\Delta\tau=0$ case, while dashed line for $\Delta\tau=0.1/g$ case.
}
\label{fidelity_CNOT}
\end{figure}
%--------------------------------------------------------------------

Next we estimate the fidelity of the process relative to the ideal case.
The process fidelity is defined by $F=\frac{1}{16}\Tr[\hat{\chi}_\text{ideal}\hat{\chi}]$ where $\hat{\chi}_\text{ideal}$ is the input-output matrix for the ideal \textsc{cnot} gate operation.
We easily find $F(r,\Delta\tau)=\frac{1}{32}\{\alpha^{(1)}_{++}+\alpha^{(2)}_{++}+2(2c_+r\alpha^{(3)}_{++}+c_+^2r\alpha^{(4)}_++c_+r^2\alpha^{(5)}_+)\}$.
Figure \ref{fidelity_CNOT} shows the process fidelity as a function of polarization $r$.
Solid and dashed lines correspond to $\Delta\tau=0$ and $\Delta\tau=0.1/g$ cases, respectively.
From this figure, we can see that a loss of process fidelity due to the fluctuation of operation time becomes larger as the polarization $r$ becomes larger.
At the perfect polarization $r=1$, the loss of fidelity is about 5\% for $\Delta\tau=0.1/g$.
Notice that in the case of $\Delta\tau=0$, the process fidelity is simply described as $F(r,0)=\frac{1}{16}(1+3r)^2$.
When the polarization of channel electrons is $r=0.6$ ($r=0.8$), we obtain the process fidelity $F=0.49$ ($F=0.72$) for $\Delta\tau=0$.

Moreover, We investigated the spin polarization $r$ required to the confirmation of entanglement creation by CNOT gate and process tomography.
When the input state is a separable state $\frac{1}{\sqrt{2}}(|{\u}\ket_X+|{\d}\ket_X)\otimes|{\u}\ket_A$ and fluctuation $\Delta\tau=0$, we have found that the output state is measured as an entangled state for the spin polarization of the electron $r>\frac{1}{\sqrt{3}}\sim 0.58$.

\section{Conclusion} \label{conclusion}
In this paper we have investigated the realization of the \textsc{cnot} gate and characterized the gate operation by quantum process tomography for a two-qubit system, realized by electrons confined in self-assembled quantum dots embedded in the spin field-effect transistor.
In this system, there exist the spin exchange interactions between the neighboring qubits and one can only measure the spin state of the edge qubit by the spin-blockade measurement.
We have shown that the \textsc{cnot} gate operation can be realized by using the spin exchange interaction and several single-qubit rotations within the coherence time of qubits.
The quantum process tomography can be also achieved in this system.
We have taken into account of the fluctuation of operation time $\Delta\tau$ and the imperfection of polarization $r$ of channel electrons as sources of decay of fidelity.
The process fidelity of \textsc{cnot} decreases at most 5\% by the fluctuation of the operation time and the values of $0.49$ and $0.72$ are obtained for the channel spin polarizations of $r=0.6$ and $0.8$, respectively.
Furthermore we have estimated the spin polarization required to the confirmation of entanglement creation by CNOT gate and process tomography.
We have found that $r>\frac{1}{\sqrt{3}}\sim0.58$.
This will serve as a target value of the necessary polarization in order to experimentally implement the proposed device.
In the recent experiments, spin-injection efficiency of 50\% has already been achieved \cite{R08} and experimental technology has been intensively advancing.
Therefore the \textsc{cnot} gate operation and process tomography for the proposed system should be able to be achieved experimentally.

%--------------- Acknowledgment ---------
\bigskip
\begin{acknowledgments}
This work is partly supported by a Grant-in-Aid for Scientific Research (C) from JSPS, Japan, by the Strategic International Research Cooperative Program from JST, and by the Program to Disseminate Tenure Tracking System and a Grant-in-Aid for Young Scientists (B) both from the MEXT, Japan.
\end{acknowledgments}

%--------------- References ------------

%%%%%%%%%%%%%%%%%%%%%%%%%%%%%%%%%%%%%%%%%%%%%%%%%%%%%%%%%%%%%%%%%%%%%%%%%%%%%

\begin{thebibliography}{30}
\bibitem{T1}
The spin relaxation and the dephasing time in a self-assembled InAs quantum dot are up to a few nanoseconds and a few hundred picoseconds, respectively, at temperature of 10 K.
See also Ref. \onlinecite{WASM05}.
\end{thebibliography}
\end{document}